\documentclass[twocolumn,           
               showpacs,            
               preprintnumbers,     
               aps,                 
               prd,                 
               a4paper,             
               groupedaddress,      
               nofootinbib,         
               tightenlines,        
               floats,floatfix      
               ]{revtex4}

\usepackage{graphicx}
\usepackage{subfigure}
\usepackage{latexsym}
\usepackage{amsmath,amssymb}        
\usepackage[draft=false]{hyperref}

\renewcommand{\(}{\left(}
\renewcommand{\)}{\right)}
\renewcommand{\[}{\left[}
\renewcommand{\]}{\right]}

\input colordvi

\begin{document}



\title{Oscillations in the inflaton potential?}
\author{C\'edric Pahud$^{1,2}$, Marc Kamionkowski$^2$, and Andrew
  R. Liddle$^1$} 
\affiliation{$^1$Astronomy Centre, University of Sussex, Brighton BN1 9QH,
     United Kingdom\\$^2$California Institute of Technology, Mail Code
     130-33, Pasadena, CA 91125}
\date{\today}
\preprint{}


\begin{abstract}
We consider a class of inflationary models with small oscillations
imprinted on an otherwise smooth inflaton potential.  These
oscillations are manifest as oscillations in the power spectrum of
primordial perturbations, which then give rise to oscillating
departures from the standard cosmic microwave background power
spectrum.  We show that current data from the Wilkinson Microwave
Anisotropy Probe constrain the amplitude of a sinusoidal variation in
the inflaton potential to have an amplitude less than $3\times
10^{-5}$.  We anticipate that the smallest detectable such
oscillations in Planck will be roughly an order of magnitude smaller,
with slight improvements possible with a post-Planck
cosmic-variance limited experiment.
\end{abstract}

\maketitle


\section{Introduction}

Cosmic microwave background (CMB) experiments continue to be
consistent with the simplest predictions of inflationary models, even
as the data become increasingly precise \cite{CMBexperiments}.
Constraints on the amplitude and spectral index of primordial density
perturbations, and on the amplitude of the inflationary
gravitational-wave background, can now be used to constrain the
parameter space of the inflaton potential.  In most of the current
analyses the inflaton potential is parameterized in terms of its
amplitude $V(\phi)$ and first and second derivatives, $V'(\phi)$ and
$V''(\phi)$.

Given the recent advances in data quality, as well as the improvements
anticipated with forthcoming experiments (e.g., the Planck satellite
\cite{planck}, to be launched within the next year), it is worth
asking whether the data can be used to study more complicated forms of
the potential.  For example, Adams et al. \cite{Adams:1997de} showed
that supergravity-inspired models may give rise to inflaton potentials
with a large number of steps.  Each step corresponds to a
symmetry-breaking phase transition in a field coupled to the inflaton.
The inflaton mass then changes suddenly when each transition occurs.
These steps are responsible for unusual inflaton dynamics, often
represented as a hybrid inflation model
\cite{Adams:2001vc}, and
they will create oscillating features in the primordial power
spectrum. Oscillations can also be directly imprinted on the inflaton
potential itself, due to some trans-Planckian physics
\cite{Brandenberger:2000wr,Easther:2002xe,Martin:2003sg}.  Finally,
some other mechanisms may create features in the primordial power
spectrum
\cite{Starobinsky:1992}.
There may also be empirical motivations to consider more complicated
potentials, as several CMB analyses suggest that the CMB power
spectrum may be better fit by primordial power spectra with features
than by smooth power spectra
\cite{Shafieloo:2003gf,Kogo:2003yb,Sealfon:2005em,Verde:2008ix}.

In this paper, we consider a class of inflationary models that feature
periodic oscillations imprinted on a smooth inflaton potential, which
then give rise to oscillations in the primordial power spectrum.  We
look for these oscillations in the Wilkinson Microwave Anisotropy
Probe (WMAP) data, and then determine the smallest oscillation
amplitude that will be probed with forthcoming experiments.  Our work
is somewhat similar to that in Ref.~\cite{Covi:2006ci}, which
considers oscillations in the primordial power spectrum that arise from
a step in the inflaton potential, but differs in that we consider
wiggles in the inflaton potential itself. There is also related work
in Ref.~\cite{Hunt:2004vt}, which considers oscillations in the CMB
power spectrum from a rapid phase transition during inflation.
However, the work most closely related to ours is that in
Ref.~\cite{Wang:2002hf}, which considers oscillations in the inflaton
potential in natural-inflation models.  While they focused on
constraints from existing data, we forecast also the detectability
with future measurements.

The plan of this paper is as follows. Section \ref{S:themodel}
introduces the model and discusses the calculation of the power
spectrum.  Section \ref{S:powerspectra} presents numerical
results for the primordial, matter, and CMB power spectra.
Section \ref{S:wmapsearch} presents results of our search for
oscillations in the WMAP data.  Section \ref{S:detectability}
discusses the forecasts for the detectability of oscillations in
future experiments; we consider here both Planck and a
post-Planck cosmic-variance limited experiment.  We summarize
and provide some concluding remarks in Section
\ref{S:conclusions}.


\section{The oscillating model}
\label{S:themodel}

In order to study the effects of small oscillations in the
inflaton potential, we begin with a simple base inflationary model
that is consistent with current data.  We consider the simplest
such model, namely a quadratic inflaton potential,
$V_0(\phi)=\frac{1}{2}m^2\phi^2$, with $m\simeq \sqrt{8 \pi}
\times 10^{-6}~M_{\mathrm{Pl}}$.  This potential has a corresponding CMB
amplitude $A_s=1.2\times10^{-9}$ for a $\Lambda$CDM model.
Once normalized with the best-fit WMAP5 data, we get $n_s=0.96$
for a pivot scale $k_\star=0.002~\mathrm{Mpc}^{-1}$, consistent
with current constraints \cite{wmap5}.

We then superimpose on this smooth potential a sinusoidal
fluctuation to give a potential of the form
\begin{equation}
     V(\phi) = \frac{1}{2} m^2\phi^2 \[1+\alpha
     \sin\(\frac{\phi}{\beta M_{\mathrm{Pl}}}+\delta\)\], 
\label{potential}
\end{equation}
parameterized by an amplitude of oscillations and a
parameter $\beta$ that characterizes the frequency. The
amplitude $\alpha$ is assumed to be small so that the inflaton
does not get stuck in one of the local minima introduced in the
potential by the oscillations; we discuss the precise
constraint later.  We choose the phase $\delta=0$ and explain
below why our results will be  the same for different values.

The homogeneous dynamics are dictated by the Friedmann equation
for the scale factor and the inflaton equation of motion.  The
Friedmann equation (in units where $8\pi G=c=\hbar=1$) for a
universe containing a scalar field $\phi(t)$ with potential
$V(\phi)$ is
\begin{equation}
     3H^2\equiv 3\left(\frac{\dot a}{a}\right)^2
     =\frac{1}{2}\dot\phi^2+V(\phi),
\end{equation}
and the scalar-field equation of motion is
\begin{equation}
     \ddot\phi+3H\dot\phi=-\frac{dV}{d\phi}.
\end{equation}
We solve these coupled differential equations for the scale
factor $a(t)$ and scalar field $\phi(t)$ numerically.
The solution is insensitive to our choice of initial conditions
for $\phi(t)$, as the solution exhibits an attractor behaviour
if the field begins high enough in the potential \cite{LL}.

We then turn our attention to the perturbations.  We express the
power spectrum ${\cal P}_{\cal R}$ of the primordial curvature
perturbation with the horizon-crossing approximation \cite{Stewart:1993bc,LL},
\begin{equation}
     {\cal P}_{\cal R}(k) = [1-2(2C+1)\epsilon_H + 2C\eta_H
     ] \(\frac{H^2}{2\pi|\dot\phi|}\)^2,
\label{pk}
\end{equation}
where $C=-2+\ln2+b\simeq-0.73$, with $b$ the Euler--Mascheroni
constant. The right-hand side is evaluated at $k=aH$,
and the Hamilton--Jacobi slow-roll parameters are
\begin{equation}
     \epsilon_H \equiv \frac{1}{2}
     \frac{\dot\phi^2}{H^2},~~~\eta_H \equiv -
     \frac{\ddot\phi}{H\dot\phi}\,.
\end{equation}

The next step is to relate a value of $\phi$ to a comoving
wavenumber $k$ that crosses the horizon at that value of
$\phi$.  To do so, we note that the number of $e$-foldings of
inflation between a time $t$ and the end of inflation is
\begin{equation}
     N(t) \equiv \mathrm{ln}\frac{a(t_{\mathrm{end}})}{a(t)},
\end{equation}
and in terms of $k$, it is
\begin{equation}
     N(k)\simeq55-\mathrm{ln}\frac{k}{a_0 H_0},
\end{equation}
where $a_0$ is the scale factor today (which we choose to be
$a_0=1$), and $H_0= (h/3000)~\mathrm{ Mpc}^{-1}$ with $h\simeq0.72$.
Note that the uncertainty around the 55 $e$-foldings of 
inflation at horizon crossing is about 5, and comes mainly 
from the uncertainty in the reheating process \cite{Liddle:2003as}.

The distance scale $\lambda$ that exits the horizon varies roughly as
the exponential of the change $\delta \phi$ in the inflaton
$\phi$. Hence an oscillation in $\phi$ in the inflaton potential
should give rise to oscillations in the logarithm of the wavenumber
$k$ in the primordial power spectrum, and thus in the logarithm of the
CMB multipole moment~$\ell$.

We then calculate the temperature and polarization (TT, EE, and TE)
spectra for the model using the {\tt CAMB} code \cite{camb},
using the current best-fit parameters for a $\Lambda$CDM model
from the WMAP 5-year results \cite{wmap5}.  We neglect the
B-mode polarization, as it comes about either through
gravitational waves or gravitational lensing and is always small
compared with the E-mode polarization.

\section{Power Spectra}
\label{S:powerspectra}

\begin{figure}
\begin{center}
\includegraphics[width=0.9 \linewidth]{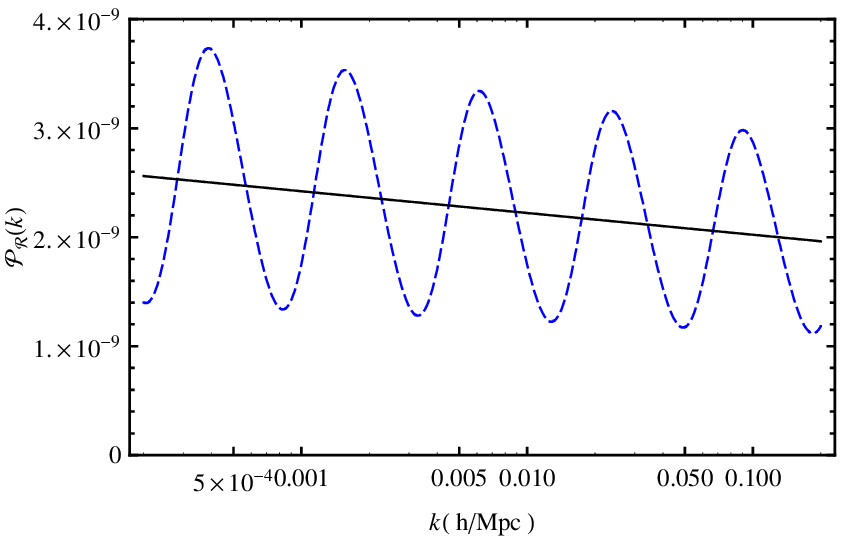}
\end{center}
\vspace{0.25cm}
\begin{center}
\includegraphics[width=0.9 \linewidth]{MatterPS.eps}
\caption{Primordial power spectrum (top) and matter power spectrum
     (bottom) of the smooth inflaton potential (solid) and
     oscillating potential (dashed). The oscillating-potential
     parameters are
     [$\alpha,\beta$]=[$5\times10^{-4},3\times10^{-2}$].  The
     amplitude is chosen to be large to clearly show the
     effect of the oscillations.}
\label{f:primordial}
\end{center}
\end{figure}

\begin{figure}
\includegraphics[width=0.9 \linewidth]{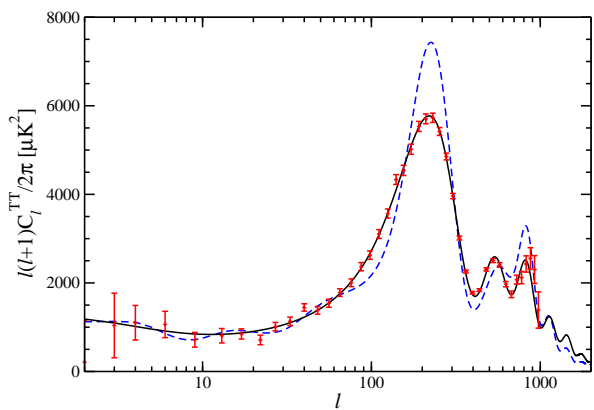}
\caption{The CMB power spectrum corresponding to the models shown
     in Fig.~\protect\ref{f:primordial}.  The WMAP5 data are
     superimposed \protect\cite{wmap5}. The error bars include
     both the cosmic variance and instrumental noise.}
\label{f:angular}
\end{figure}

We begin by showing in Fig.~\ref{f:primordial} the primordial
power spectrum, as well as the present matter power spectrum, for the
standard smooth potential and for an oscillating potential with
parameters chosen to be
[$\alpha,\beta$]=[$5\times10^{-4},3\times10^{-2}$]. The
oscillations in the power spectra are nearly 
sinusoidal in $\log k$ with almost constant amplitude.
The corresponding CMB angular power spectrum, shown in
Fig.~\ref{f:angular}, reflects this behaviour. The mapping from
the three-dimensional matter power spectrum to the
two-dimensional CMB power spectrum slightly smooths the
wiggles.  In this example, the frequency $\beta$ of the
inflationary potential's oscillations has been chosen so that
one of the primordial oscillation peaks lines up with the first
acoustic peak. However, this coincidence will not be generic.

\section{A Search in WMAP data}
\label{S:wmapsearch}

\begin{figure}[t]
\includegraphics[width=0.9 \linewidth]{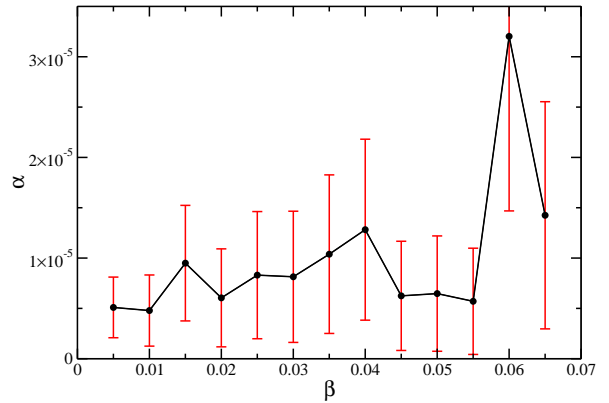}
\caption{The best-fit amplitude $\alpha$ and its standard error,
     considering WMAP5 data, for our range of frequencies $\beta$.}
\label{f:alpha_MCMC}
\end{figure}

Given the CMB predictions discussed above, it is straightforward to
search the existing CMB data from WMAP for these oscillations, and we
have carried out a Markov Chain Monte Carlo (MCMC) analysis to do so.
Fig.~\ref{f:alpha_MCMC} presents the results of this analysis,
assuming that all other cosmological parameters are known. For each
point, we run four Markov chains.  The results provide an indication
of the best-fit amplitude $\alpha$ as a function of the assumed
frequency $\beta$. These amplitudes are comparable to their standard
errors, and so we conclude that there is no evidence for oscillations
in the WMAP data.  Instead, we infer an upper limit $\alpha\lesssim
3\times 10^{-5}$ from WMAP.  We note that a corresponding MCMC
analysis has already been performed in Ref.~\cite{Covi:2006ci} for the
WMAP3 data for a similar potential. They find a precision on the
amplitude of oscillations of the order of $10^{-5}$ for WMAP3,
consistent with our results.

\section{Detectability}
\label{S:detectability}

We now estimate the smallest oscillation amplitude $\alpha$ that
will be detectable with future experiments. To do so, we first
consider for simplicity the temperature power spectrum only. We
suppose that each multipole moment $\ell$ can be
measured with a standard error
\begin{equation}
     \sigma_l =\sqrt{\frac{2}{(2l+1)f_{\rm sky}}} \,
     \(C_l+N_l\),
\end{equation}
where
\begin{equation}
     N_l \equiv \left(\overline{\Delta T}^2 \theta_b^2 \right)
     \exp \frac{l^2\theta_b^2}{8\ln2}\,
\end{equation}
is the contribution from the detector noise, $\overline{\Delta T}$ is the
detector noise per angular-resolution element, and $\theta_b$ is
the beam width.  We then estimate the error on $\alpha$ by
\cite{Jungman:1995bz}
\begin{equation}
     \(\frac{1}{\sigma_\alpha}\)^2 = \sum_l \(\frac{\partial
     C_l}{\partial \alpha}\)^2 \frac{1}{\sigma_l^2 (\alpha)},
\label{Fisher}
\end{equation}
and we choose an amplitude $\alpha=5\times10^{-5}$ to compute the $C_l$ derivatives.
This estimate assumes that all other cosmological parameters are
known, and as such, provides an optimistic estimate.  However, the
true value, obtained by marginalizing over all other cosmological
parameters, will probably not be too much worse, as there are no
strong degeneracies between these oscillations and any other
cosmological parameters. The acoustic oscillations oscillate in
$\ell$, while these oscillate in $\log \ell$, and so they should
not be strongly degenerate.

In order to improve our results, we then include 
the polarization and temperature-polarization power spectra as well.
In that case, the generalization of the expression for the smallest 
detectable oscillation takes the form \cite{Kamionkowski:1996ks}
\begin{equation}
     \(\frac{1}{\sigma_\alpha}\)^2 = \sum_l \sum_{{\rm A,A'}} {
     \partial C_l^{\rm 
     A} \over \partial \alpha} \left[\Psi^{-1}\right]_{AA'}  {
     \partial C_l^{\rm 
     A'} \over \partial \alpha},
\end{equation}
for A $=$ TT, EE, TE where $[\Psi^{-1}]_{AA'}$ are elements of the
inverse of $\Psi$, the covariance matrix; its elements are given
in Ref.~\cite{Kamionkowski:1996ks}.

\begin{figure}[t]
\includegraphics[width=0.9 \linewidth]{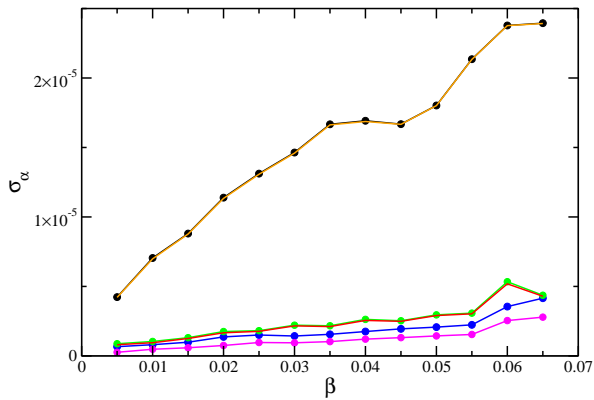}
\caption{From the top to the bottom, by pairs, the smallest 
     detectable amplitude $\sigma_\alpha$ as a function of $\beta$, 
     for WMAP, Planck and a cosmic-variance limited experiment.
     The upper part of each pair includes the temperature data only, 
     and the lower one the polarization data as well. 
     Note that the temperature and
     temperature--polarization curves for WMAP and Planck are
     effectively degenerate and thus appear to be one curve.}
\label{f:alpharesults_a}
\end{figure}

\begin{figure}[t]
\includegraphics[width=0.9 \linewidth]{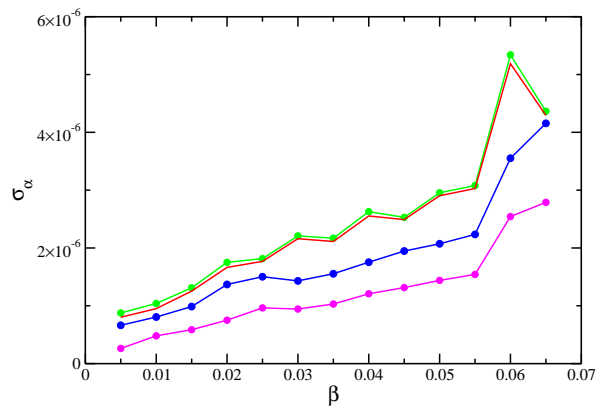}
\caption{Magnification of Fig.~\ref{f:alpharesults_a} for Planck
and the cosmic-variance limited experiment.}
\label{f:alpharesults_b}
\end{figure}

In Figs.~\ref{f:alpharesults_a} and \ref{f:alpharesults_b}, we
plot the smallest detectable amplitude $\sigma_\alpha$, from
these analyses, as a function of $\beta$, for WMAP (in
Fig.~\ref{f:alpharesults_a} only),  Planck, and a hypothetical 
cosmic-variance limited experiment.  For WMAP, we simply
consider the measured uncertainties from the 5-year results
\cite{wmap5}.  For Planck, the forecast is done considering the three most 
sensitive temperature channels, of specifications similar to the
HFI channels of frequency 100 GHz, 143 GHz, and 217 GHz \cite{planck}. The
intensity sensitivities (detector noise) of these channels are
taken as 6.8 $\mu$K, 6.0 $\mu$K, and 13.1 $\mu$K respectively,
corresponding to the values quoted for two complete sky surveys.
These are average sensitivities per pixel, where a pixel is a
square whose side is the FWHM extent of the beam. The FWHMs (beam
widths) of these channels are given as 9.5 arcmin, 7.1 arcmin, and
5.0 arcmin, respectively. The composite noise spectrum for
the three temperature channels is obtained by inverse-variance
weighting the noise of individual channels
\cite{Knox:1995dq,Cooray:1999kg}. For polarization we take only
one channel, the 143 GHz channel, of sensitivity 11.5 $\mu$K,
and FWHM 7.1 arcmin. Finally, for the cosmic-variance limited
experiment,  the fractional sky covered is taken to be
0.8 for all $\ell$, and we use simulated data out to an
$\ell_{{\rm max}}$ of 2000, as for Planck.

The statistical analysis presented in this Section is intended
to forecast the uncertainty on parameters measured
by future experiments.  It is nevertheless re-assuring that this
forecast, when applied to WMAP, recovers roughly the limits
obtained from the detailed MCMC analysis.
Fig.~\ref{f:alpharesults_a} illustrates the advantage of
Planck over WMAP for an oscillation search.  While the smallest
detectable amplitude is $O(10^{-5})$ for WMAP, it is
$O(10^{-6}$) for Planck.  Looking at Planck more closely, in
Fig.~\ref{f:alpharesults_b}, we find a smooth variation of
$\sigma_\alpha$ except at some frequencies, such as $\beta=0.03$
and $\beta=0.04$ for instance. These deviations arise
from correlations between the acoustic peaks and the
primordial peaks, coming from the inflaton potential, in the CMB
power spectrum. When varying the frequency, the bumps thus
created are aligned differently with the acoustic
oscillations. For example, the choice of $\beta=0.03$ considered
in Fig.~\ref{f:angular} is particular as many bumps are well
aligned with the acoustic peaks. This effect is however much
less important at high frequency, as our results suggest. 

The temperature-only and temperature--polarization curves
coincide for both WMAP and Planck.  This follows because the
polarization amplitude is much smaller and neither WMAP nor Planck
will measure polarization to the cosmic-variance limit.  The
contribution to the total signal-to-noise from polarization in
these experiments is thus small.  The degeneracy
is broken in the cosmic-variance limited experiment.  In this
case, the precisions with which the temperature and polarization
power spectra can be measured are roughly the same, resulting in
roughly a $\sqrt{2}$ improvement to the oscillation sensitivity.

Figs.~\ref{f:alpharesults_a} and \ref{f:alpharesults_b} indicate that a
higher frequency of oscillations (i.e.\ a smaller $\beta$) allows a
smaller detectable oscillation amplitude. This can be understood by
considering Eq.~\eqref{Fisher}, whose result is mainly governed by the
difference between the chaotic and the oscillating curve, in terms of
amplitudes. For a given amplitude of oscillations in the CMB power
spectrum, but different frequencies, the difference between these two
curves is almost the same (it would be exactly the same if the slope of
the chaotic spectrum was constant). However, in our analysis, we are
considering a given amplitude $\alpha$ in the inflaton potential itself.
When the frequency of oscillations in the inflaton potential increases,
so does that in the CMB power spectrum, but the corresponding amplitude
in the CMB spectrum also increases (rather than staying constant). As
such, for a given $\alpha$, the difference between the chaotic and
oscillating curve increases when $\beta$ decreases, and hence
$\sigma_{\alpha}$ decreases.

We can now also justify our neglect of the phase $\delta$ introduced
in Section \ref{S:themodel}.  Its effect would simply be to shift the
deviations on the smooth curves.  In Section \ref{S:themodel}, we
assumed that $\alpha$ was sufficiently small that inflation was not
interrupted, and we can now justify that assumption.  The constraint
on $\alpha$ from WMAP is already $\alpha \lesssim 3\times 10^{-5}$.
In the case of the most critical values considered in our analysis,
being [$\alpha,\beta$]=[$5\times10^{-5},5\times10^{-3}$], we get
[$\epsilon_H^{\rm{max}},| \dot\epsilon_H^{\rm{max}} |$] $\simeq$
[$1\times10^{-2},2\times10^{-7}$] and [$| \eta_H^{\rm{max}} |,|
\dot\eta_H^{\rm{max}} |$] $\simeq$ [$2\times10^{-1},2\times10^{-4}$],
well within the slow-roll regime.

\section{Conclusions}
\label{S:conclusions}

The purpose of our paper has been to study whether features in
the inflaton potential, beyond the standard slow-roll
parameterization, can be detected with current and forthcoming
CMB experiments.  To do so, we have considered the possibility
that the inflaton potential varies sinusoidally with the
inflaton.  A sinusoidal variation of $V(\phi)$ with $\phi$
induces a variation in the matter and CMB power spectra that is
sinusoidal in $\log k$ and $\log \ell$, respectively.  

Our analysis of the WMAP data indicates that the upper bound on
$\alpha$, the oscillation amplitude, is roughly $3\times 10^{-5}$.  To
derive this bound, we argued that the degeneracy between $\alpha$ and
other cosmological parameters is weak, and then assumed in the data
analysis that the other parameters were fixed.  We anticipate that a
more complete analysis, including marginalization over other
parameters, will weaken this bound, but only slightly.  We then showed
that the sensitivity of Planck to the oscillation amplitude will be
greater, relative to WMAP, by roughly an order of magnitude.  For both
Planck and WMAP, the constraint comes primarily from the temperature.
Planck might be improved upon slightly by a cosmic-variance limited
experiment. The advantage of such an experiment is that the additional
information in the polarization may then be fully capitalized upon.

On the theory side, we
used the horizon-crossing approximation to calculate the power
spectrum.  The work of Ref.~\cite{Wang:2002hf} suggests that
numerical integration of the perturbation equations will result
in a slight suppression of the oscillation amplitude, relative
to that we have obtained, but certainly by no more than a factor
of two.  If so, then our bounds may be accordingly weakened.  We
leave investigation of these data-analysis and theory questions
for future research.

\begin{acknowledgments}

C.P.\ acknowledges the hospitality of the Moore Center for
Theoretical Cosmology and Physics at Caltech.  C.P.\ was
supported in part by the Swiss Sunburst Fund and the Royal
Astronomical Society, M.K.\ by DoE DE-FG03-92-ER40701 and
the Gordon and Betty Moore Foundation, and A.R.L.\ by STFC (UK).
\end{acknowledgments}


\end{document}